\documentclass[]{JHEP3}
\usepackage{graphicx}
\usepackage{bm}

\usepackage{subfigure, epsfig, rotate, epsf,setspace, amssymb, amsxtra,amsfonts}

\newcommand{\average}[1]{\big{\langle} #1 \big{\rangle}}
\newcommand{\prop}{\mathcal{G}}
\newcommand{\propdag}{\prop^\dagger}

\newcommand{\zedtwo}{\mathbb{Z}(2)}
\newcommand{\zedtwocross}{\mathbb{Z}(2)\otimes\mathbb{Z}(2)}
\newcommand{\propf}[3]{\prop_{#1}(#2\leftarrow #3)}
\newcommand{\propfdag}[3]{\propdag_{#1}(#2\rightarrow #3)}

\newcommand{\fry}[1]{\vec{y},#1}
\newcommand{\frx}[1]{\vec{x},#1}
\newcommand{\frz}[1]{\vec{z},#1}

\newcommand{\frv}[1]{\vec{v},#1}
\newcommand{\fru}[1]{\vec{u},#1}

\newcommand\edinb{SUPA, School of Physics, The University of Edinburgh\\Edinburgh EH9 3JZ, UK}

\newcommand\mainz{Institut f\"{u}r Kernphysik, Johannes Gutenberg-Universit\"{a}t Mainz\\D-55099 Mainz, Germany}

\title{Use of stochastic sources for the lattice determination of light quark physics}
\author{P.A.~Boyle, C.~Kelly, R.D.~Kenway\\\edinb\\Email: \email{paboyle@ph.ed.ac.uk}, \email{christopher.kelly@ed.ac.uk}, \email{rdk@epcc.ed.ac.uk}}
\author{A. J\"{u}ttner\\\mainz\\Email: \email{juettner@kph.uni-mainz.de}}

\author{RBC \& UKQCD collaboration}

\preprint{Edinburgh 2008/17, MKPH-T-08-06}
\abstract{
In this paper we investigate the benefits of using $\zedtwocross$ single timeslice stochastic sources for the calculation of light quark physics on the lattice. Meson 2-point correlators measured using sources stochastic in only spin and those stochastic in both spin and colour indices are compared to point source correlators on the unit gauge and on a $16^3\times 32$ Domain Wall QCD ensemble. It is found that the use of stochastic sources gives a considerable improvement in statistics for the same computational cost. The neutral kaon mixing matrix element $B_K$ is also calculated on this ensemble with stochastic sources, but we conclude that the stochastic method offers no significant advantage over the traditional gauge-fixed wall source approach which already offers an exact volume average. We also discuss the application to semileptonic form factors in conjunction with partially twisted boundary conditions.}

\begin{document}
\section{Introduction}
\label{section:introduction}
Meson correlation functions are fundamental to phenomenological applications of lattice QCD. Pseudoscalar states in particular are used to determine quark masses, the LECs of the Chiral Effective Lagrangian, and CKM relevant observables such as $B_K$ and the $K_{l3}$ form factor.

In this paper we discuss two and three-point functions using the interpolating operator $\mathcal{O}_{1,2}$
\begin{equation}\mathcal{O}_{1,2}=\bar{\psi}_1 \Gamma \psi_2\label{eqn:mesinterp}\end{equation}
and its conjugate to create and annihilate mesonic states. Here $\Gamma$ is a product of gamma matrices set to give the operator the correct quantum numbers, and $\psi_i$ are quark fields of flavour $i$. 

After Wick contraction, the correlation functions contain a product of spin matrices with quark propagators $\mathcal{G}$. The propagators obey $\gamma^5$-hermiticity
\begin{equation}\prop_i(\vec{x}, \tau \leftarrow \vec{y}, t) = \gamma^5\prop_i^\dagger(\vec{x}, \tau \rightarrow \vec{y}, t)\gamma^5\,,\end{equation}
for which the arrow indicates the direction of quark flow, and it is understood that the source indices of the conjugate propagator lie on the left and those of the unconjugated propagator on the right. Even on a lattice of modest size, the propagator matrix is extremely large and thus only a subset can be calculated within a reasonable timescale. This is performed by solving the matrix equation
\begin{align}
\psi(\vec{y},t) &\equiv \sum_{\vec{x},\tau}\mathcal{M}^{-1}(\vec{y},t\;;\;\vec{x},\tau)\eta(\vec{x},\tau)\\
                &=  \sum_{\vec{x},\tau}\prop(\vec{y},t\leftarrow \vec{x},\tau) \eta(\vec{x},\tau)
\end{align}
for the `solution' vector $\psi$, where $\eta$ is a complex vector `source' occupying some region of space, $\mathcal{M}$ is the Dirac matrix, and the matrices are contracted over spin and colour indices. This equation can be solved using, for example, the iterative conjugate gradient algorithm.

Typically the point source is used, consisting of unit spin and colour vectors on a single space-time point $(\vec{x}_0, t_0)$. The $12$ possible spin and colour source vectors are usually written as a unit spin and colour matrix, forming a matrix source $\tilde{\eta}$
\begin{equation}
\begin{array}{rl|l}
\tilde{\eta}(\vec{x},t)  & = \mathbb{I}_{4\times 4}\otimes \mathbb{I}_{3\times 3} &  (\vec{x},t) = (\vec{x}_0, t_0)\\
	         & = 0 & \mathrm{otherwise}
\end{array}\,,
\end{equation}
where $\mathbb{I}_{N\times N}$ is the $N\times N$ unit matrix. 

Solutions evaluated from these sources are matrices consisting of the subset of elements of the propagator from a single space-time point to all other points on the lattice, for all combinations of spin and colour indices at source and sink, thus requiring $12$ inversions of the Dirac matrix. These solutions are typically referred to as one-to-all propagators.

Use of a localised source increases the sensitivity of the measurements to local fluctuations in the gauge fields; for example propagators evaluated from regions in which the Dirac matrix has a localised near zero mode will produce large outliers in the results. The statistical distribution of observables should be much better behaved if a volume average is included in each measurement.

This paper is concerned with stochastic vector sources, for which the elements of the source are randomly drawn from a distribution $\mathcal{D}$ that is symmetric about zero. A set of $N_{\mathrm{hits}}$ randomly generated lattice volume filling sources
\begin{equation}
\{\eta^{(n)}(x)_{a\alpha} \in \mathcal{D}|n=1\ldots N_{\mathrm{hits}} \}\,,
\end{equation}
referred to as a set of `hits' of the stochastic source, for which $a$ is a colour index and $\alpha$ a spin index, has the property that, in the limit of $N_{\mathrm{hits}}\rightarrow \infty$
\begin{equation}\langle\eta^{(n)}_{a\alpha}(x)\eta^{\dagger(n)}_{b\beta}(y)\rangle_n \equiv \frac{1}{N} \sum_{n=1}^{N_{\mathrm{hits}}} \eta^{(n)}_{a\alpha}(x)\eta^{\dagger(n)}_{b\beta}(y) \rightarrow \delta_{x,y}\delta_{ab}\delta_{\alpha\beta}\,. \label{eqn:stochproperties}\end{equation}
Lattice volume stochastic sources have been used in the past in order to estimate the entire propagator matrix \cite{z2-all-to-all,physlett-b389,9810021,all-to-all,trinlat-dilute,0603007}. Here the solutions are referred to as stochastic all-to-all propagators. Dong and Liu \cite{z2-all-to-all} demonstrated that sources with $\zedtwo$ noise
$\mathcal{D}= \zedtwo = \{+1,-1\}$ or generally $\mathcal{D} = \mathbb{Z}(N)$ for any $N$, deviate less from the orthonormality condition of equation (\ref{eqn:stochproperties}) for a fixed number of hits than those estimated with Gaussian or `double-hump' Gaussian-like distributions. Foster and Michael \cite{9810021} suggest that the optimal choice is the c-number distribution $\mathcal{D}=\zedtwocross$, which contains random $\zedtwo$ numbers in both its real and imaginary parts, i.e. 
\begin{equation}\mathcal{D} = \Big{\{}\frac{1}{\sqrt{2}}(\pm 1 \pm i)\Big{\}}\,.\end{equation}

Stochastic all-to-all propagators are generally very noisy, and thus are often abandoned in favour of the traditional one-to-all propagators apart from situations in which they are necessary \cite{all-to-all,trinlat-dilute}, such as when the number of gauge configurations is limited and one must extract as much information as possible from each.

This paper details an exploration into an alternate use of stochastic sources for the calculation of meson correlators at zero momentum, based upon the work of Foster and Michael \cite{9810021} (appendix), a form of which is referred to as the `one-end trick' \cite{0603007}. This method has been used by the ETM collaboration  for the calculation of meson two-point \cite{Boucaud:2007uk,Blossier:2007vv,Boucaud:2008xu} and three-point functions \cite{Simula:2007fa}. The aim of this paper is to determine whether these meson correlators can be calculated more cheaply and with better statistics than the traditional point source, and to investigate the competitiveness of extensions of this method to a range of matrix elements compared to the respective traditional approaches.

The layout of the paper is as follows. The method of the one-end trick is introduced in the context of the two-point correlation functions, based upon Foster and Michael's \cite{9810021} description, followed by the details of the two stochastic source types we have chosen to overcome the highlighted issues. The stochastic two-point correlators are compared to point source correlators on the unit gauge in order to prove the correct behaviour of the stochastic correlators on a trivial gauge. Results for pseudoscalar and vector correlators, analysed on a $16^3\times 32$ RBC/UKQCD Domain Wall QCD ensemble, are then presented and discussed. We then assess the performance of the stochastic source technique in the computation of the neutral kaon bag parameter $B_K$. To this end we compare the evaluation of the relevant three-point functions with a single stochastic wall source fixed to Coulomb gauge and a naive gauge fixed wall source. We also compare the single-wall approach to the two-wall approach of Antonio et al. \cite{Antonio:2007pb}. We finish with a discussion of the calculation of hadronic form factors using the stochastic method, as adopted by the ETM collaboration \cite{Simula:2007fa} and the RBC \& UKQCD collaboration \cite{Boyle:2008yd}.

\section{Two-point correlation functions}
\label{section:twopoint}

Using the interpolating operator $\mathcal{O}_{1,2}$ of equation (\ref{eqn:mesinterp}), the two-point meson correlator is defined as
\begin{equation}C(t^\prime;\vec{p}) = \sum_{\vec{x},\vec{y}} e^{-i\vec{p}\cdot(\vec{y}-\vec{x})} \langle\mathcal{O}_{2,1} (\vec{y},t)\mathcal{O}_{1,2}(\vec{x}, \tau)\rangle\,,\end{equation}
where $t^\prime \equiv t - \tau$ and $\vec{p}$ is a momentum. Performing the Wick contraction and neglecting the disconnected contribution that appears when $\psi_1=\psi_2$, we find
 
\begin{equation}C(t^\prime;\vec{p}) = \sum_{\vec{x},\vec{y}} e^{-i\vec{p}\cdot(\vec{y}-\vec{x})}
\times\;\mathrm{tr}\left(\gamma^5\Gamma\prop_1(\vec{y},t \leftarrow \vec{x}, \tau)
 \Gamma\gamma^5 \prop^\dagger_2(\vec{x}, \tau \rightarrow \vec{y}, t)\right)\,,\label{eqn:correlator-definition}\end{equation}
where the trace is over spin and colour indices and $\prop_i$ is a quark propagator of flavour $i$. 

\subsection{The One-End Trick}
\label{section:oneend}
Consider the meson two-point correlator of equation (\ref{eqn:correlator-definition}) at zero momentum and insert a product of Kronecker delta functions

\begin{equation}
C(t^\prime;\vec{0})= \sum_{\vec{x},\vec{y},\vec{z}}
(\gamma^5\Gamma)_{\alpha\beta}\propf{1\;b\beta,c\kappa}{\fry{t}}{\frx{\tau}}\Big{[}\delta_{\kappa\lambda}\delta_{cd}\delta_{\vec{x},\vec{z}}\Big{]}(\Gamma \gamma^5)_{\lambda\rho} \propfdag{2\;d\rho,b\alpha}{\frz{\tau}}{\fry{t}}\,.
\label{eqn:oneendpartway}
\end{equation}
Here Greek letters represent spin indices and Roman letters colour indices. Using stochastic `wall' sources 
\begin{equation}
\begin{array}{rl|l}
\eta^{(n)}_{c\kappa}(\frx{t}|\tau) & \in \mathcal{D} & t=\tau\\
& = 0 & t\neq \tau
\end{array}\label{eqn:sources}
\end{equation}
we can replace the delta functions by a hit average
\begin{equation}
\delta_{\kappa\lambda}\delta_{cd}\delta_{\vec{x},\vec{z}} = \langle\eta^{(n)}_{c\kappa}(\frx{\tau}|\tau)\eta^{\dagger(n)}_{d\lambda}(\frz{\tau}|\tau)\rangle_n\,,\label{eqn:deltas}
\end{equation}
using the orthonormality condition of equation (\ref{eqn:stochproperties}). Inserting this into equation (\ref{eqn:oneendpartway}) we find that the correlator becomes the scalar-product of two solution vectors, one of which is dependent on the matrix $\Gamma$:
\begin{equation}
C(t^\prime;\vec{0})= \sum_{\vec{y}}\Big{<}\left(\gamma^5\Gamma\psi_1^{(n)}(\fry{t}|\tau)\right)\,\cdot\,\psi^{\Gamma\:\dagger(n)}_2(\fry{t}|\tau)\Big{>}_n\,.\label{eqn:oneendsol}
\end{equation}
Here
\begin{equation}
\psi^{(n)}_1(\fry{t}|\tau) \equiv \sum_{\vec{x}}\propf{1}{\fry{t}}{\frx{\tau}}\;\eta^{(n)}(\frx{\tau}|\tau)\,,\\
\label{eqn:oneendsolutions1}\end{equation}
\begin{equation}
\psi^{\Gamma\:(n)}_2(\fry{t}|\tau) \equiv \sum_{\vec{x}}\propf{2}{\fry{t}}{\frx{\tau}}(\Gamma \gamma^5)^\dagger \eta^{(n)}(\frx{\tau}|\tau)\,
\label{eqn:oneendsolutions2}
\end{equation}
and the vectors $\gamma^5\Gamma\psi$ and $\psi^{\Gamma\:\dagger}$ are contracted at the sink location. The source indices are contracted automatically by the stochastic average, completing the trace in the $N_{\mathrm{hits}}\rightarrow \infty$ limit.

The advantages of the above method for the calculation of the entire meson spectrum are reduced by the necessity to calculate $\psi^{\Gamma}$ for each of the $16$ $\Gamma$ matrices, requiring $16$ inversions per stochastic hit. This can be reduced to $4$ inversions per hit by calculating the spin structure explicitly. We refer to these as spin-explicit or \textit{SEM} sources following ref. \cite{Viehoff:1997wi}. These sources are further discussed in section \ref{z2sem-subsection}.

\subsection{Pseudoscalar \textit{Z2PSWall} source}
\label{section:stochsources}
\label{section:z2pswall}
For the pseudoscalar case $\Gamma=\gamma^5$, the solution vectors $\psi_i$ and $\psi_i^\Gamma$ are identical, allowing for calculation of the pseudoscalar meson correlator with only a single inversion of the Dirac matrix per hit and valence mass. This completely stochastic source type, defined as per equation (\ref{eqn:sources}), will henceforth be referred to as a \textit{Z2PSWall} source.

Let us consider the structure of the \textit{Z2PSWall} source in more detail. The spin and colour space components can be represented as a single $12$-component column vector $\Xi^{(n)}(\vec{x})$, such that the source has the form
\begin{equation}
\eta^{(n)}(\vec{x},t|\tau) = \{\delta_{t,\tau}\} \otimes \Xi^{(n)}(\vec{x})\,.
\end{equation}
The elements of $\Xi^{(n)}(\vec{x})$ are stochastically sampled from the chosen distribution
\begin{equation}
\{\Xi^{(n)}_{i}(\vec{x})\in \mathcal{D} |i=1\ldots 12\}
\end{equation}
and thus the vectors obey the orthonormality condition
\begin{equation}
M(\vec{x},\vec{y})\equiv \average{\Xi^{(n)}(\vec{x})\otimes\Xi^{(n)\;\dagger}(\vec{y})}_n \rightarrow \delta_{\vec{x},\vec{y}}\;\mathbb{I}_{12\times 12}\,,
\label{eqn:stochmatrixprod}
\end{equation}
where $\otimes$ is the vector direct product
\begin{equation}M_{ij}(\vec{x},\vec{y}) \equiv \average{\Xi^{(n)}_i(\vec{x})\Xi^{*\;(n)}_j(\vec{y})}_n\; \big{|} i,j=1\ldots 12\,.\label{eqn:estimator2}\end{equation}
From this we see that $M$ is Hermitian
\begin{equation}M^\dagger_{ij}(\vec{x},\vec{y})=M^*_{ji}(\vec{y},\vec{x})=M_{ij}(\vec{x},\vec{y})\,.\end{equation}
By equation (\ref{eqn:estimator2}) we see that the diagonal elements of $M$ (i.e. those with both $\vec x = \vec y$ and $i=j$) are unity. Subtracting these elements, we define the stochastic noise matrix $K$ as
\begin{equation}
K(\vec{x},\vec{z})= M(\vec{x},\vec{z})- \delta_{\vec{x},\vec{z}}\;\mathbb{I}_{12\times 12}\,.
\end{equation}
Thus, we can split the correlator of equation (\ref{eqn:oneendsol}) into signal and noise components
\begin{equation}
 \begin{array}{rl}C(t^\prime)&\equiv \sum_{\vec{y}}\langle \psi_1^{(n)}(\fry{t}|\tau)\psi^{\dagger(n)}_2(\fry{t}|\tau)\rangle_n\\\\
& = C_S(t^\prime) + \Delta C(t^\prime)
\end{array}\,,\label{stochcorr}
\end{equation}
where
\begin{equation}
C_S(t^\prime)\equiv \sum_{\vec{x},\vec{y}}\mathrm{tr}\left(\propf{1}{\fry{t}}{\frx{\tau}}\propfdag{2}{\frx{\tau}}{\fry{t}}\right)
\end{equation}
is the gauge-invariant signal component and
\begin{equation}
\Delta C(t^\prime) = \sum_{\vec{x},\vec{y},\vec{z}}\mathrm{tr}\left(\propf{1}{\fry{t}}{\frx{\tau}}K(\vec{x},\vec{z})\propfdag{2}{\frz{\tau}}{\fry{t}}\right)
 \label{stochnoisecorr}
\end{equation}
is the noise component. Here the trace over sink indices has been reintroduced as this is now a product over matrices. The noise component contains a mixture of gauge-invariant and gauge-dependent pieces for finite $N_{\mathrm{hits}}$ as well as contributions from other meson correlators. This can be seen by decomposing $K(\vec{x},\vec{y})\in \mathbb{C}^{144}$ onto the basis $\{\lambda_r\otimes \Gamma_i\}$, composed of the 8 Gell-Mann matrices $\{\lambda_r\;| r=1\ldots 8\}$, the $3\times 3$ unit matrix $\lambda_0 = \mathbb{I}_{3\times 3}$, the $4\times 4$ unit matrix $\Gamma_0=\mathbb{I}_{4\times 4}$, and the 15 tensor combinations of the gamma matrices $\{\Gamma_i\;|i=1\ldots 15\}$. The components of this basis are orthogonal under the trace operation
\begin{equation}\mathrm{tr}\left(\lambda_r\otimes \Gamma_i\;\;\lambda_s\otimes \Gamma_j\right) = \alpha_r\delta_{rs}\delta_{ij}\,,\end{equation}
where $\alpha_0=12\;,\;\alpha_r=8\:|r\neq 0$. Under this decomposition
\begin{equation}
K(\vec{x},\vec{y}) = \sum_{i,r}A^i_r(\vec{x},\vec{z})\;\lambda_r \otimes \Gamma_i\,,
\end{equation}
with c-number coefficients $A^i_r(\vec{x},\vec{z})$. Applying this decomposition to the noise component $\Delta C(t^\prime)$, we find
\begin{equation}
\Delta C(t^\prime)= \sum_{i,r}\sum_{\vec{x},\vec{y},\vec{z}}A^i_r(\vec{x},\vec{z})\;\mathrm{tr}\left(\propf{1}{\fry{t}}{\frx{\tau}}\;\lambda_r \otimes \Gamma_i\;\propfdag{2}{\frz{\tau}}{\fry{t}}\right)\,.
\end{equation}
With reference to equation (\ref{eqn:correlator-definition}), we see that for all components bar the unit matrix contribution $\lambda_0\otimes\Gamma_0$,  the spin and colour matrices at the source location are different from those at the sink. Therefore these components are formed from the Green's function of two \textit{different} interpolating operators: the pseudoscalar $\mathcal{O}(\lambda_0\otimes\gamma^5)$ at the sink with the polluting `unwanted operators' $\mathcal{O}(\lambda_r\otimes\Gamma_i\gamma^5)$ at the source. The contaminating noise is small in the case of the scalar, vector and tensor Dirac structures as the pseudoscalar is the lightest state. These contributions are eliminated in the ensemble average due to parity, and also in the $N_{\mathrm{hits}}\rightarrow \infty$ limit. The overlap with the axial state $A_0$ is eliminated in the hit limit and is empirically smaller in magnitude than the pseudoscalar signal. We shall introduce spin-explicit sources for use with non-pseudoscalar measurements. The effects of the gauge-dependent terms with $\vec{x}\neq \vec{z}$, which we refer to as `cross-terms', and components with $\lambda_r \neq \mathbb{I}_{3\times 3}$ are discussed further in section \ref{section:averaging}.

The \textit{Z2PSWall} source can be implemented within a software framework designed for $12\times 12$ matrix sources, such as the point source discussed in section \ref{section:introduction}, allowing for the reuse of existing propagator contraction code without further modification. This can be achieved by placing the stochastic source vector $\Xi^{(n)}(\vec{x})$ on the first column of an empty $12\times 12$ matrix $\Phi(\vec{x},t)$ on each lattice site of the wall
\begin{equation}
 \begin{array}{rl|l}\Phi^{(n)}(\vec{x},t) & = \left(\begin{array}{ccccc}: & 0 & 0 & 0 & \cdots\\
								\Xi^{(n)}(\vec{x}) & 0 & 0 & 0 & \cdots\\
								: & : & : & : & \cdots
                                            
                                  \end{array}\right) & t=\tau\\
                                  & = 0 & t\neq\tau
                \end{array}\,.
\end{equation}
The solution $\psi^{\prime\:(n)}(\vec{y},t|\tau)$ is matrix valued
\begin{equation}
\psi^{\prime\:(n)}(\vec{y},t|\tau)_{AC} \equiv \sum_{\vec{x}} \mathcal{M}^{-1}_{AB}(\vec{y},t\;;\;\vec{x},\tau)\Phi^{(n)}_{BC}(\vec{x},\tau)\,,
\end{equation}
but with all columns zero bar the first ($C=0$). Here $A,B,C$ are spin-colour indices. Our inverter, of course, checks for a zero norm source vector before inversion. With this implementation, the direct product that forms the stochastic matrix $M$ of equation (\ref{eqn:stochmatrixprod}) simplifies to the stochastic average of the matrix product
\begin{equation}
M(\vec{x},\vec{y}) = \langle\Phi(\vec{x})\Phi^\dagger(\vec{y})\rangle_n\,,  
\end{equation}
such that the meson two-point function of equation (\ref{eqn:oneendsol}) becomes simply a trace over a product of matrices
\begin{equation}
C(t^\prime;\vec{0}) = \sum_{\vec{y}}\mathrm{tr}\;\langle\psi_1^{(n)}(\vec{y},t|\tau)\psi_2^{(n)\;\dagger}(\vec{y},t|\tau)\rangle_n\,,
\end{equation}
which has the same form as the standard point source meson correlator contraction.

\subsection{Spin-explicit \textit{Z2SEMWall} sources}
\label{z2sem-subsection}
As the \textit{Z2PSWall} source can only be used for pseudoscalar correlators, we require a source type of more general use. We can stochastically estimate the general meson two-point correlator of equation (\ref{eqn:oneendsol}) with four inversions by calculating the spin structure explicitly, using stochastic noise in colour space only. These sources were used by Viehoff et al. \cite{Viehoff:1997wi} for the calculation of the matrix element of the axial vector current between proton states, and later by Boucaud et al. \cite{Boucaud:2007uk,Boucaud:2008xu} for meson correlation functions, where they are referred to as `linked sources'.

Similarly to the point source, the four spin vectors can be combined into a single $4\times 4$ unit spin matrix. A different stochastic colour vector $\xi^{(n)}$ is used on every site of the timeslice allowing us to retain the spatial and colour delta functions in the hit average, while an explicit Kronecker delta is used for the spin components. The source has the structure
\begin{equation}
\eta^{(n)}(\vec{x},t|\tau) = \{\delta_{t,\tau}\} \otimes \mathbb{I}_{4\times 4}\otimes \xi^{(n)}(\vec{x})\,,
\end{equation}
where
\begin{equation}
\{\xi^{(n)}_{a}(\vec{x})\in \mathcal{D} |a=1\ldots 3\}
\end{equation}
obeys the orthonormality condition
\begin{equation}
M(\vec{x},\vec{y})\equiv \average{\xi^{(n)}(\vec{x})\otimes\xi^{(n)\;\dagger}(\vec{y})}_n \rightarrow \delta_{\vec{x},\vec{y}}\;\mathbb{I}_{3\times 3}\,.
\end{equation}
For a finite number of hits, the matrix $M$ can again be decomposed onto the basis of Gell-Mann matrices and the unit matrix $\lambda_0$. As before we expect that the gauge dependent terms will be suppressed by the ensemble average.

As with the \textit{Z2PSWall} source, the \textit{Z2SEMWall} can be placed within a $12\times 12$ matrix, allowing for the reuse of existing measurement code:
\begin{equation}
 \Phi(\vec{x},t) = \left(\begin{array}{cccc}
                   1 & 0 & 0 & 0\\
		   0 & 1 & 0 & 0\\
		   0 & 0 & 1 & 0\\
		   0 & 0 & 0 & 1 
                   \end{array}\right) \otimes 
		   \left(\begin{array}{ccccc} 
		   :                 & 0 & 0\\
		  \xi^{(n)}(\vec{x}) & 0 & 0\\
		   :                 & 0 & 0
                    \end{array}\right)
		    \end{equation}
at $t=\tau$ and zero elsewhere.

\subsection{Hit averaging and the ensemble average}
\label{section:averaging}
The elements of the matrix $K$ are either stochastically generated or zero. Therefore, provided the distribution $\mathcal{D}$ is symmetric about zero, the combined probability distribution of the gauge configurations $U$ and the stochastic matrix $K$ has the property
\begin{equation}
P[U, K]=P[U]P[K]=P[U]P[-K]\,.
\end{equation}
Here we consider the noise components of the correlator containing gauge-dependent terms. Terms that break gauge invariance are naturally suppressed by averaging over gauge-equivalent configurations within an ensemble. However the Monte Carlo sampling of a gauge orbit is typically quite slow as one increases the ensemble size, due to autocorrelations between configurations. The stochastic method improves upon this by explicitly removing these terms through the symmetric fluctuations of $K$ about zero.

If sufficiently many hits per configuration are sampled then the cancellation will be near exact, while for a smaller number of hits this will take place stochastically as the distribution of gauge fields and sources is jointly sampled. For a large enough ensemble the difference between having few hits and having many is likely to be small. 

\subsection{Demonstration on the unit gauge}
\label{section:unitgauge_meson}
Following Foster and Michael \cite{9810021}, we use $\mathcal{D}=\zedtwocross$ noise. We perform our initial analysis on the unit gauge configuration, for which all gauge links are unity, in the Domain Wall QCD framework. Due to the translational invariance of this gauge field, the point source solution is exactly equal to the volume averaged propagator, and therefore this configuration is ideal for the demonstration of the convergence of the stochastic correlators to the exact volume averaged solution as the number of hits is increased.

In order to demonstrate the convergence as a function of the number of hits, we calculated the pseudoscalar meson two-point correlator with a valence quark mass of 0.04 in lattice units using different numbers of hits of both \textit{Z2PSWall} and \textit{Z2SEMWall} sources. For each choice of the number of hits $N_{\mathrm{hits}}$, we repeat the process 80 times and estimate the standard error on the mean of these 80 correlation functions, each comprising of $N_{\mathrm{hits}}$ solutions. In order to avoid having to generate new data for each $N_{\mathrm{hits}}$, the stochastic estimates used for each of the 80 measurements were randomly drawn from a large pool of single-hit stochastic estimates of the correlation function, ensuring that for a given $N_{\mathrm{hits}}$ no data point is drawn more than once. This provides us with 80 independent measurements for each $N_{\mathrm{hits}}$ and minimises correlation between values of $N_{\mathrm{hits}}$.

Figure \ref{figure:t16_unit} shows a plot of the means and standard errors of these distributions using the correlator at $t=16$. For the pseudoscalar two-point function it appears that the \textit{Z2PSWall} is converging better, especially given that it requires one quarter of the number of inversions needed for the \textit{Z2SEMWall} source. However, as described in section \ref{z2sem-subsection}, we find that the \textit{Z2SEMWall} benefits from its exact spin structure in the full calculation.

\FIGURE[t]{
\includegraphics[scale=0.4, angle=270]{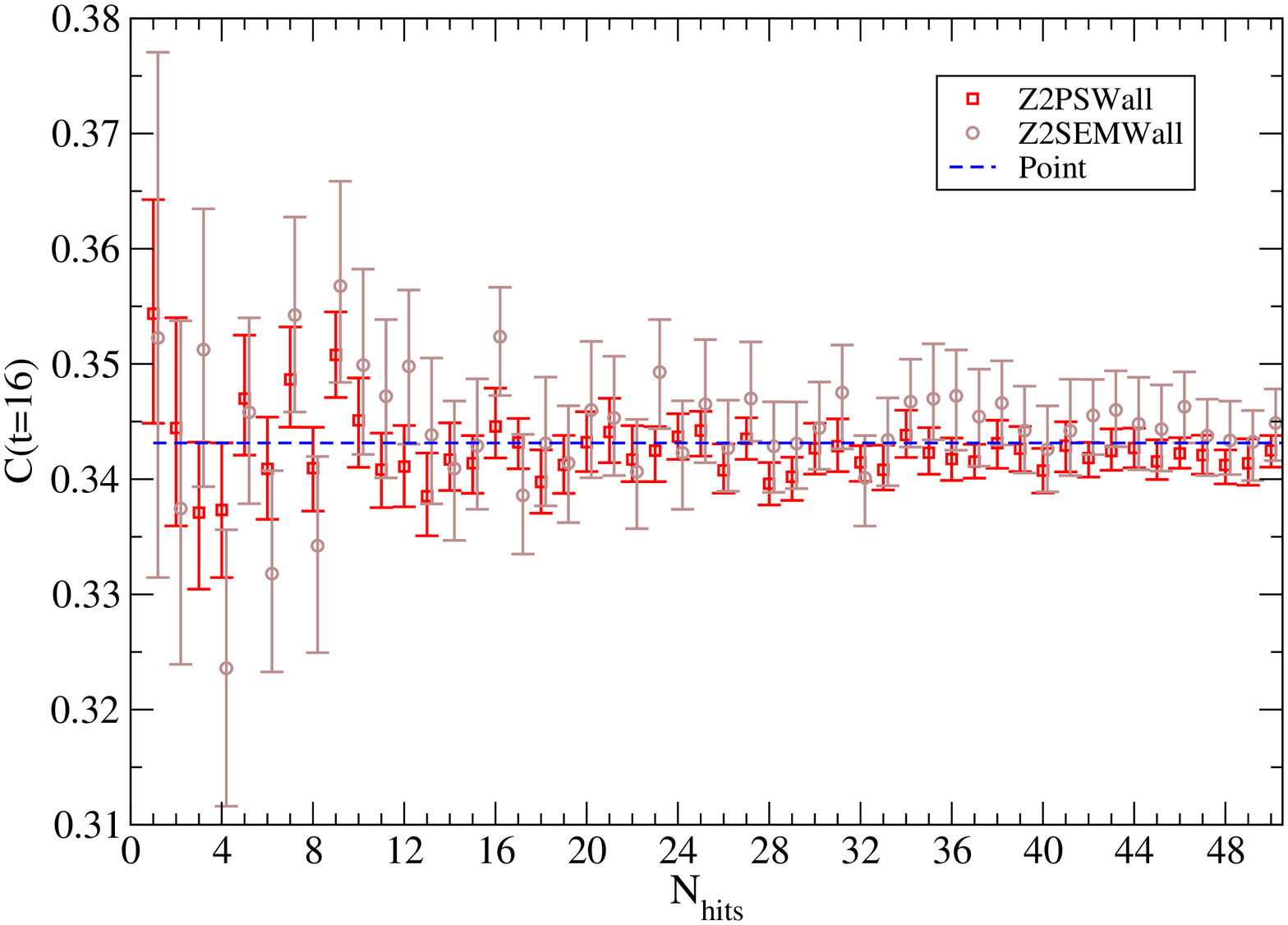} 

\caption{Demonstration of the dependence of the trivial gauge pseudoscalar meson correlator on the number of stochastic hits $N_{\mathrm{hits}}$ of \textit{Z2PSWall} and \textit{Z2SEMWall}. These are compared to the point source correlator which is the exact solution for this gauge configuration. We use 80 separate measurements, each consisting of the average of $N_{\mathrm{hits}}$ stochastic estimates in order to determine the error. For each $N_{\mathrm{hits}}$ the stochastic estimates are randomly drawn from the pool of estimates in order to minimise correlations between the data points. We ensure that each data point is not drawn more than once per $N_{\mathrm{hits}}$ such that the measurements are independent.}
\label{figure:t16_unit}
}

 \subsection{Two-point meson correlator results}
 \label{section:results_meson}
\TABLE[ht]{
\centering
\begin{tabular}{l|l}
\hline\hline
Lattice size & $16^3\times 32$\\
Action & Domain Wall\\
Gauge Action &Iwasaki\\
Domain wall height & $1.8$\\
$L_s$ & 16\\
$\beta$ & $2.13$\\
$\frac{1}{a}$ (GeV)& 1.73(3)\\
Sea quark masses (latt. units)& $m_{u} = 0.01$, $m_s = 0.04$\\
\hline\hline
\end{tabular}
\caption{Ensemble properties.}
\label{table:ensembleparams}
}

\FIGURE[p]{	
\includegraphics[scale=0.4, angle=270]{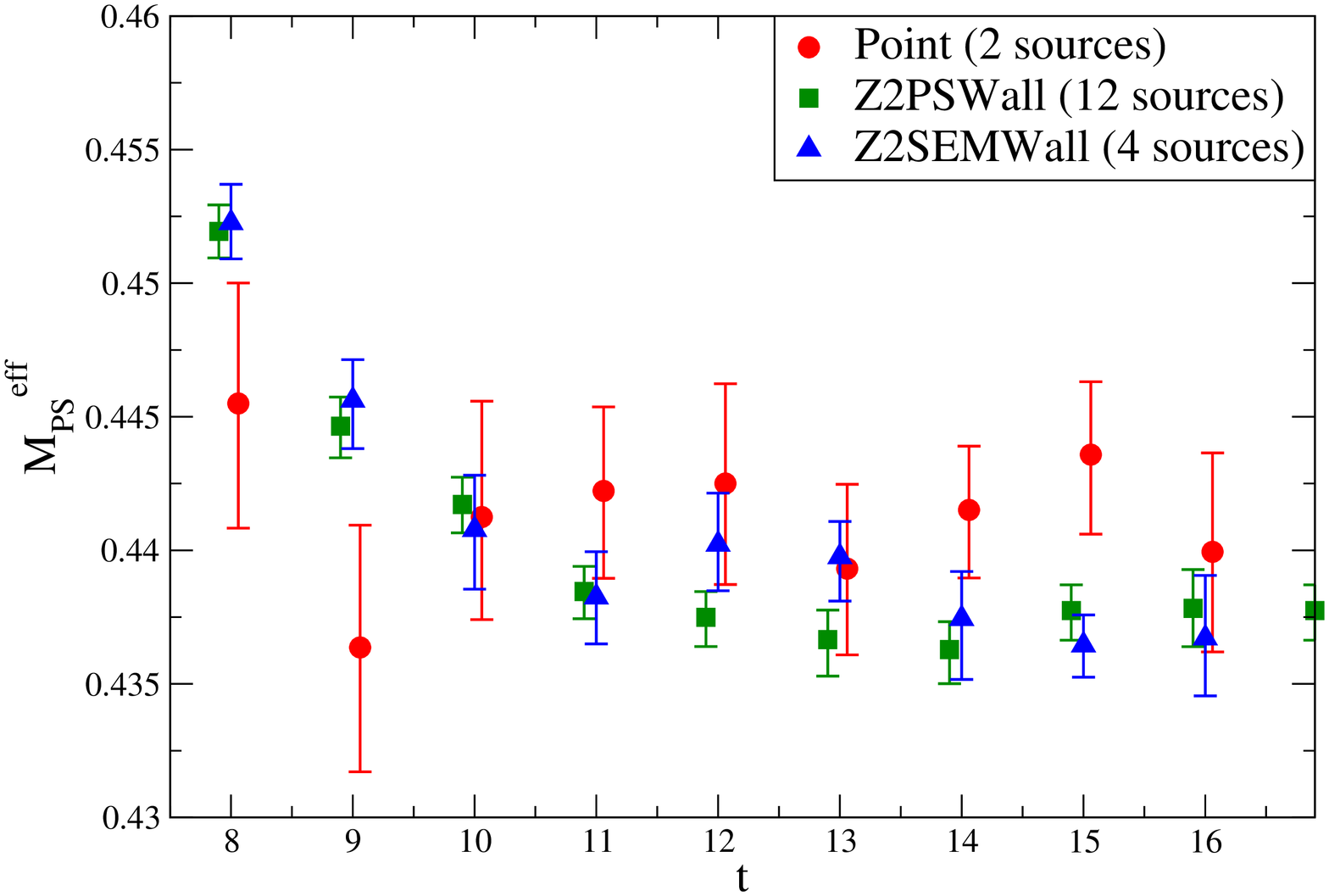}  
\caption{Pseudoscalar meson effective mass plot from averaged correlators with a bin size of 8 configurations. This is not a cost comparison. The points have been slightly shifted for clarity.}
\label{figure:avgps_meff}
} 
\clearpage

In order to take the study further, we formed the pseudoscalar meson two-point function on 392 configurations (separated by 5 molecular dynamics time-steps) of an RBC-UKQCD $16^3\times 32$ 2+1 flavour Domain Wall QCD ensemble with properties detailed in table \ref{table:ensembleparams}, for which previous results are available for comparison \cite{ed-dwf-07}. On each configuration the propagators were calculated from \textit{Z2PSWall} sources with one hit on 12 different source timeplanes. For $\tau_{\mathrm{src}} = 0$ we included 3 further hits. In addition we generated \textit{Z2SEMWall} propagators from 4 timeplanes and point source propagators from two source origins. The propagators were calculated using the conjugate gradient algorithm with a residual of $10^{-7}$ and a valence quark mass of $0.04$ in lattice units. This large valence quark mass was chosen as it is cheaper to invert; thus allowing for better statistics for a given computational cost.

In order to take account of autocorrelations in molecular dynamics time, we binned over adjacent configurations. Based upon our analysis (appendix \ref{appendixa}) and the previous analysis \cite{ed-dwf-07} we chose a bin size of 40 molecular dynamics time units (8 configurations).

Figure \ref{figure:avgps_meff} is an effective mass plot of the averaged correlators with a bin size of 8 configurations. As before we exclude three of the four \textit{Z2PSWall} hits on timeslice zero. From this figure we chose a constant fit range of $10-16$ for point source correlators and $11-16$ for \textit{Z2PSWall} and \textit{Z2SEMWall} correlators.

Table \ref{table:psmassfits} contains the results for the pseudoscalar meson mass fits for the various sources over all 392 available configurations, where the correlators have been averaged about the central timeslice (folded) for better statistics, using the forwards-backwards symmetry of the correlator. For some choices of origin or $\tau_{\rm src}$ the correlation functions show deviations from the expected time-dependence which manifests itself in a large value for $\chi^2/{\rm d.o.f}$. These effects however disappear after averaging the correlation functions over source positions.

The \textit{Z2PSWall} fitted masses appear to be consistently lower than those of the point sources, differing by $5\sigma$ between the 12-source-averaged \textit{Z2PSWall} correlator and the point source average. This discrepancy is likely to be caused by statistical fluctuations in the point source correlators: The 12-source-averaged \textit{Z2PSWall} pseudoscalar meson result of 0.4372(9) is in much better agreement than the 2 point source averaged value of 0.4418(12) with the mass obtained in the previous analysis of 0.438(3) \cite{ed-dwf-07}. The central value and error estimates of this previous result were obtained by averaging over the pseudoscalar-pseudoscalar and axial-axial correlators for several point source smearings and locations and the error  was scaled by a factor of 1.5 to account for fluctuations in the gauge fields, and as such should not be compared unfavourably with our result.

\FIGURE[t]{
\includegraphics[scale=0.4,angle=270]{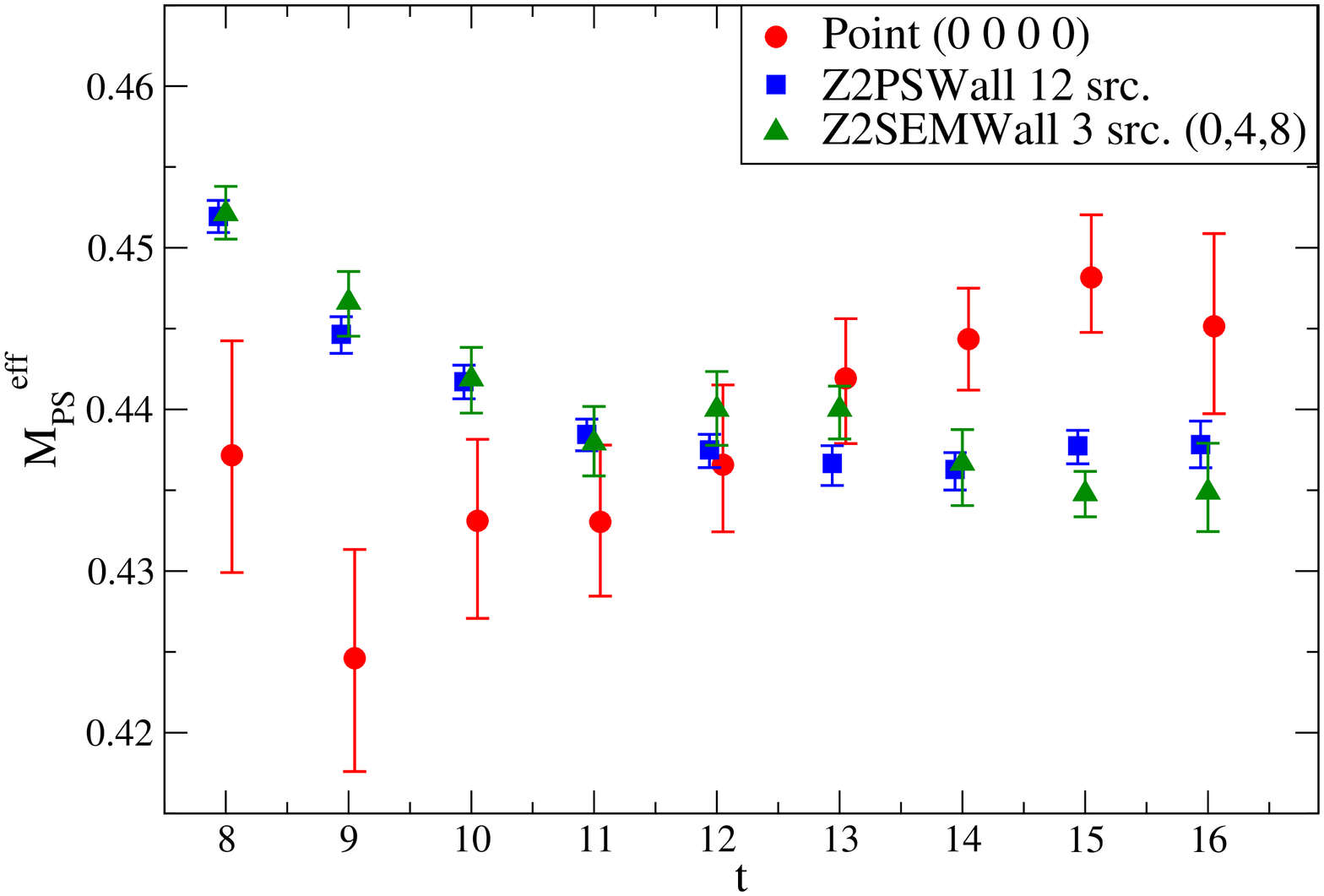}  
\caption{Pseudoscalar effective mass plots at a fixed cost of 4704 inversions of the Dirac matrix.}
\label{figure:psmeffcost}
}

\TABLE[p]{
\centering
\begin{tabular}{ccclcll}
\hline\hline
\multicolumn{7}{c}{\textbf{Point}}\\
\hline
$N_{\mathrm{src}}$ & Cost & \multicolumn{3}{l}{Origin(s)} & \multicolumn{1}{c}{Mass} & $\chi^2 /\mathrm{d.o.f.}$\\
\hline
1 & 4704 & \multicolumn{3}{l}{$x_1\equiv(0,0,0,0)$} & 0.4413(+19)(-16) & 0.543\\
1 & 4704 & \multicolumn{3}{l}{$x_2\equiv(8,8,8,16)$}& 0.4416(+20)(-23) & 2.046\\
2 & 9408 & \multicolumn{3}{l}{$x_1$ , $x_2$}   & 0.4418(+12)(-12) & 0.163
\vspace{2mm}\\
\hline
\multicolumn{7}{c}{\textbf{\textit{Z2PSWall}}}\\
\hline
$N_{\mathrm{src}}$ & Cost & $N_\tau$ & $\tau_{\mathrm{src}}$ & $N_{\mathrm{hits}}/N_{\tau}$ &  \multicolumn{1}{c}{Mass} & $\chi^2 /\mathrm{d.o.f.}$\\
\hline
1 & 392 &1 & 0  & 1 (a) & 0.4398(+19)(-16) & 0.236\\
1 & 392 &1 & 0  & 1 (b) & 0.4375(+23)(-24) & 0.300\\
1 & 392 &1 & 0  & 1 (c) & 0.4397(+24)(-24) & 0.241\\
1 & 392 &1 & 0  & 1 (d) & 0.4405(+21)(-19) & 0.481\\
1 & 392 &1 & 2  & 1 & 0.4386(+19)(-21) & 0.361\\
1 & 392 &1 & 4  & 1 & 0.4345(+24)(-26) & 0.216\\
1 & 392 &1 & 6  & 1 & 0.4323(+19)(-21) & 1.917\\
1 & 392 &1 & 8  & 1 & 0.4356(+19)(-23) & 0.286\\
1 & 392 &1 & 10 & 1 & 0.4407(+20)(-23) & 0.267\\
1 & 392 &1 & 12 & 1 & 0.4394(+21)(-22) & 0.120\\
1 & 392 &1 & 14 & 1 & 0.4397(+21)(-20) & 0.177\\
1 & 392 &1 & 16 & 1 & 0.4354(+22)(-23) & 0.069\\
1 & 392 &1 & 18 & 1 & 0.4362(+21)(-20) & 0.034\\
1 & 392 &1 & 20 & 1 & 0.4334(+20)(-21) & 0.222\\
1 & 392 &1 & 22 & 1 & 0.4390(+24)(-27) & 0.731\\
4 & 1568 & 4 & 0,4,6,8 & 1 & 0.4374(+10)(-11) & 0.632\\
4 & 1568 & 1 & 0 & 4 & 0.4393(+16)(-16) & 0.371\\
12 & 4704 & 12 & 0-22; even & 1 & 0.4372(+8)(-9) & 0.388\\
\vspace{2mm}\\
\hline
\multicolumn{7}{c}{\textbf{\textit{Z2SEMWall}}}\\
\hline
$N_{\mathrm{src}}$ & Cost & $N_\tau$ & $\tau_{\mathrm{src}}$ & $N_{\mathrm{hits}}/N_{\tau}$ &  \multicolumn{1}{c}{Mass} & $\chi^2 /\mathrm{d.o.f.}$\\ \hline
1 & 1568 & 1 & 0  & 1 & 0.4395(+15)(-14) & 0.190\\
1 & 1568 & 1 & 4  & 1 & 0.4375(+25)(-24) & 0.507\\
1 & 1568 & 1 & 8  & 1 & 0.4348(+16)(-18) & 0.707\\
1 & 1568 & 1 & 12 & 1 & 0.4406(+17)(-16) & 0.054\\
3 & 4704 & 3 & 0,4,8  & 1 & 0.4372(+9)(-11) & 0.657\\
3 & 4704 & 3 & 0,4,12 & 1 & 0.4394(+11)(-12) & 0.068\\
3 & 4704 & 3 & 0,8,12 & 1  & 0.4383(+9)(-10) & 0.127\\
3 & 4704 & 3 & 4,8,12 & 1  & 0.4375(+11)(-12) & 0.644\\
4 & 6272 & 4 & 0,4,8,12 & 1 & 0.4381(+9)(-10) & 0.378\\
\hline\hline
\end{tabular}
%}
\caption{Pseudoscalar meson mass fits for the various sources, fitting to range $10-16$ (point) or $11-16$ (stoch. sources), with a bin size of 8 configurations over an ensemble of 392 configurations. $N_\mathrm{src}$ is the total number of sources used in the fit, with the equivalent cost in inversions of the Dirac matrix detailed in the next column. The third column of the stochastic source tables contains the number of source timeslices used $N_\tau$; the fourth a list of these times $\tau_{\mathrm{src}}$; and the fifth the number of hits (stochastic samples) on those timeslices ($N_{\mathrm{hits}}/N_{\tau}$). The four independent hits of \textit{Z2PSWall} are distinguished by a Roman letter.}
\label{table:psmassfits}
}

\clearpage 

At a fixed cost of 12 inversions per configuration (4704 inversions in total), it is evident that the 12-source averaged \textit{Z2PSWall} result shows at least a factor of 2 improvement in the statistical error over the point source. The 3-source averaged \textit{Z2SEMWall} results agree with the \textit{Z2PSWall} result, and also show a consistent factor of 2 improvement in errors over the point source at the same cost.

The third- and second-to-last lines in the \textit{Z2PSWall} section of table \ref{table:psmassfits} allow for a fixed cost comparison between the use of four stochastic hits upon a single source timeslice and a single hit on four different timeslices. The 40\% reduction in the statistical error suggests that one should preferentially choose a new three-volume sample of the gauge field when forming a new stochastic source, separated in space-time and molecular dynamics time in order to maximise decorrelation between the samples.

Figure \ref{figure:psmeffcost} plots the effective mass as a function of time at the cost of 4704 inversions. It is evident that both the \textit{Z2PSWall} and \textit{Z2SEMWall} source types give significantly better plateaus than a single point source. The plateaus for the stochastic sources appear to be very similar. We believe this displays a spectacular improvement for pseudoscalar masses at no additional cost. Based upon these data we conclude that the difference in the quality of the stochastic source results at fixed cost is not large enough to warrant using \textit{Z2PSWall} sources over the spin-explicit \textit{Z2SEMWall} sources which can be used for a larger number of measurements.

In order to estimate the effectiveness of the \textit{Z2SEMWall} sources for other measurements, we consider the vector meson correlator with interpolating operator $\mathcal{O}_{1,2}= \bar\psi_1\gamma^\mu\psi_2$. Figure \ref{figure:avgvec_meff} shows the vector meson effective mass for the 4 combined \textit{Z2SEMWall} sources and the 2 point sources, where we have averaged over the three spatial gamma matrix correlators and folded about the central timeslice for better statistics. Based upon this plot we chose a fit range of 9-16 for both source types. This is not a fixed cost comparison.

Table \ref{table:vmassfits} shows the results of the fits to the vector meson correlator. As before, at fixed cost we see a reduction in error by a factor of around 2 over the point source values. Also, as with the pseudoscalar two-point function, the plateau of the vector meson effective mass (figure \ref{figure:vec_meff_cost}) for the stochastic source type is noticeably better than that of the point source.

\FIGURE[p]{
\includegraphics[scale=0.4, angle=270]{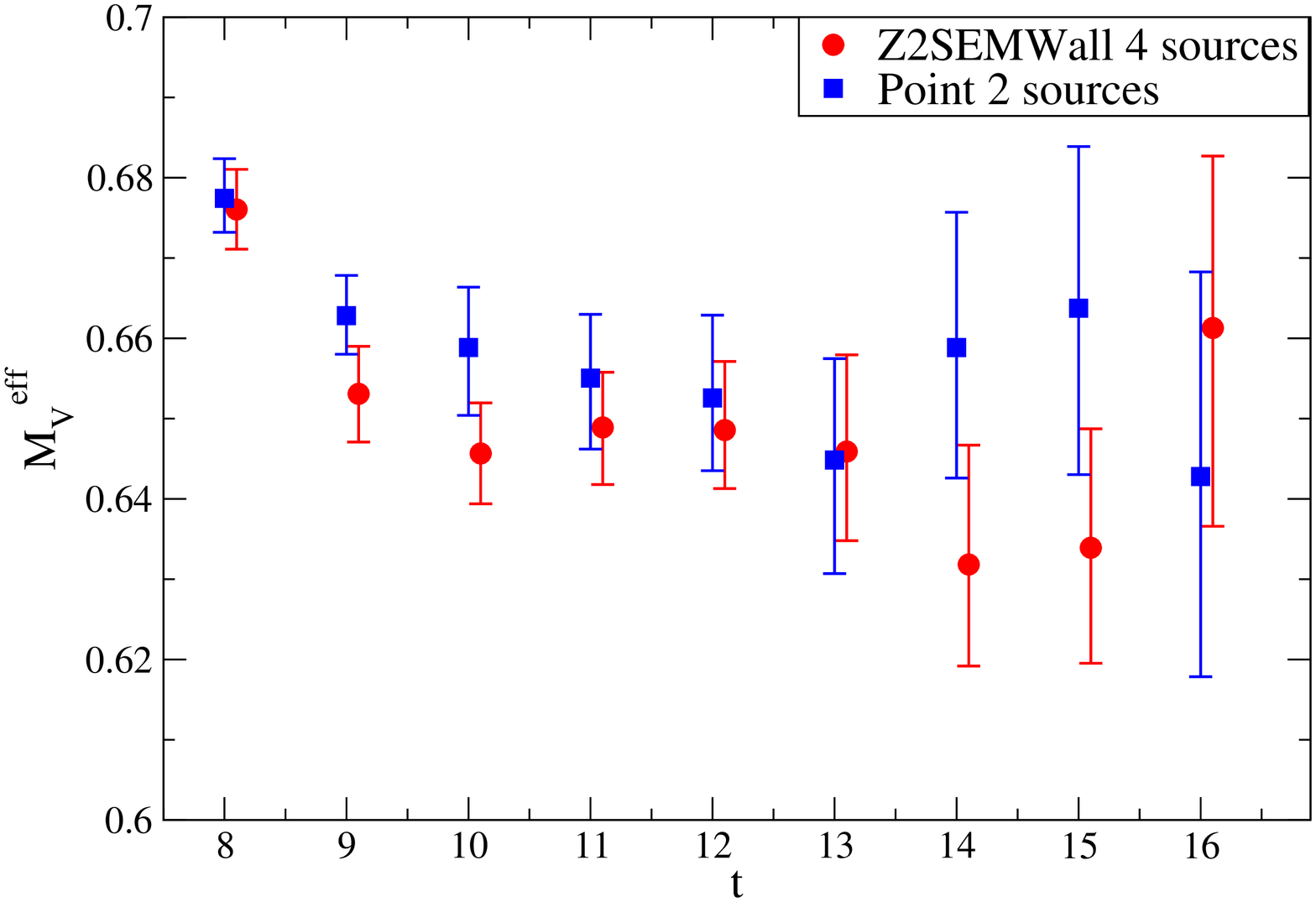}  
\caption{Vector meson effective mass plot from averaged \textit{Z2SEMWall} and point source correlators with a bin size of 8 configurations. This is not a fixed cost comparison, but can be used to select the fit ranges.}
\label{figure:avgvec_meff}
}
\FIGURE[p]{
\includegraphics[scale=0.4, angle=270]{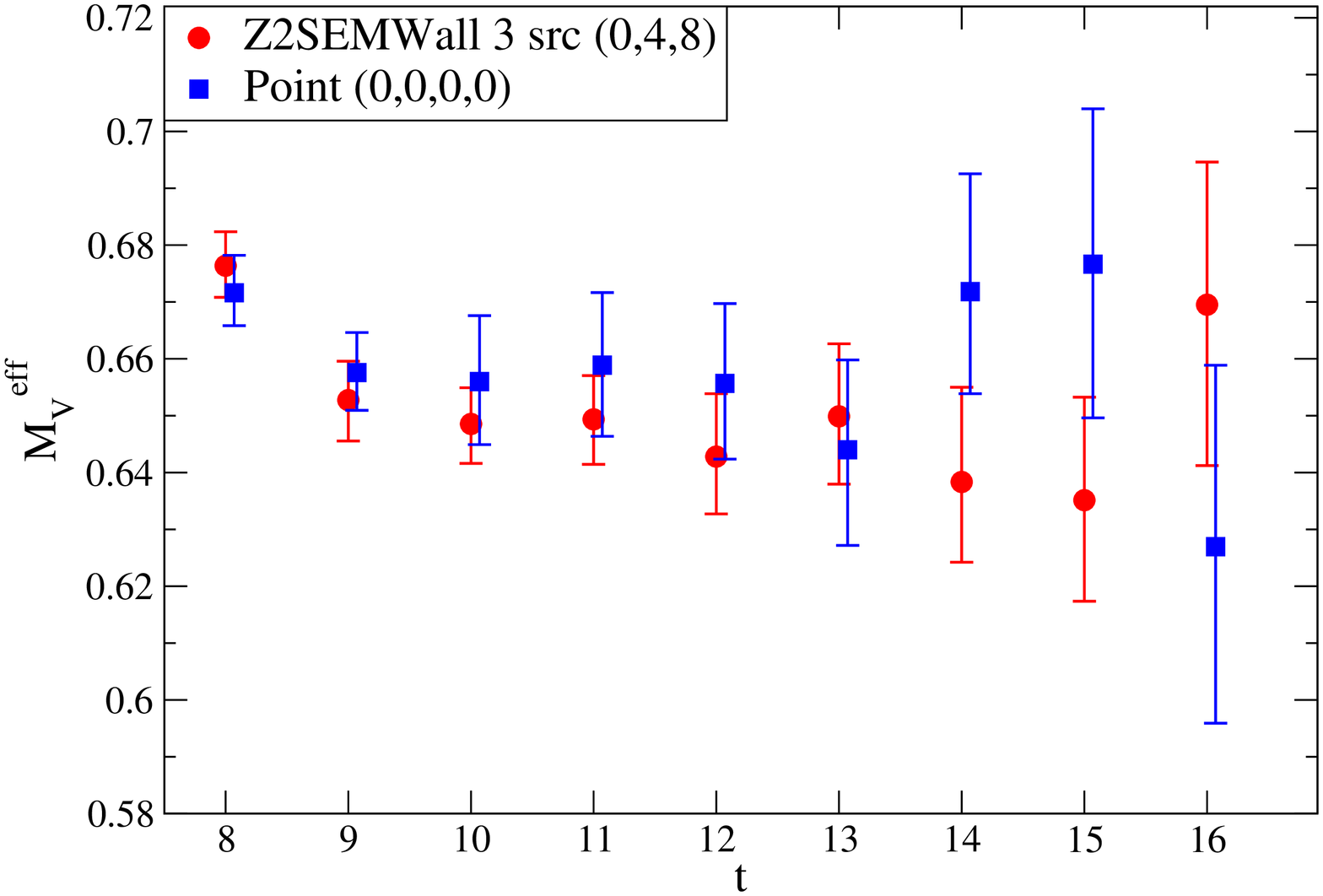}  
\caption{Vector meson effective mass plots at a fixed cost of 4704 inversions of the Dirac matrix.}
\label{figure:vec_meff_cost}
}
\clearpage

\TABLE[p]{
\centering
\begin{tabular}{ccclcll}
\hline\hline
\multicolumn{7}{c}{\textbf{Point}}\\
\hline
$N_{\mathrm{src}}$ & Cost & \multicolumn{3}{l}{Origin(s)} & \multicolumn{1}{c}{Mass} & $\chi^2 /\mathrm{d.o.f.}$\\
\hline
1 & 4704 & \multicolumn{3}{l}{$x_1\equiv(0,0,0,0)$}  & 0.656(+10)(-9) & 0.404\\
1 & 4704 & \multicolumn{3}{l}{$x_2\equiv(8,8,8,16)$} & 0.657(+8)(-9) & 0.172\\
2 & 9408 & \multicolumn{3}{l}{$x_1$ , $x_2$}  & 0.657(+7)(-7) & 0.372
\vspace{2mm}\\
\hline
\multicolumn{7}{c}{\textbf{\textit{Z2SEMWall}}}\\
\hline
$N_{\mathrm{src}}$ & Cost & $N_\tau$ & $\tau_{\mathrm{src}}$ & $N_{\mathrm{hits}}/N_{\tau}$ &  \multicolumn{1}{c}{Mass} & $\chi^2 /\mathrm{d.o.f.}$\\ \hline
1 & 1568 & 1 & 0 & 1 & 0.642(+8)(-8) & 0.207\\
1 & 1568 & 1 & 4 & 1 & 0.660(+10)(-11) & 0.328\\
1 & 1568 & 1 & 8 & 1 & 0.642(+10)(-9) & 0.407\\
1 & 1568 & 1 & 12 & 1 & 0.637(+9)(-9) & 0.538\\
3 & 4704 & 3 & 0,4,8 & 1 & 0.649(+5)(-5) & 0.153\\
3 & 4704 & 3 & 0,4,12 & 1 & 0.647(+5)(-6) & 0.101\\
3 & 4704 & 3 & 0,8,12 & 1 & 0.641(+5)(-5) & 0.377\\
3 & 4704 & 3 & 4,8,12 & 1 & 0.646(+6)(-6) & 0.457\\
4 & 6272 & 4 & 0,4,8,12 & 1 & 0.646(+4)(-5) & 0.179\\
\hline\hline
\end{tabular}
\caption{Vector meson mass fits for the various sources, fitting to range $9-16$ with a bin size of 8 configurations. Here we have used the conventions established in table \ref{table:psmassfits}.}
\label{table:vmassfits}
}
\clearpage

\section{Stochastic calculation of the $B_K$ three-point function}
\label{section:stochasticbk}
The success of the stochastic method for two-point functions motivates us to consider extending the technique to include meson
three-point functions. In this paper we consider the kaon bag parameter $B_K$, which describes the mixing between the neutral $K_0$ and $\bar K_0$ mesons:
\begin{equation}B_K^{\mathrm{latt}} \equiv \frac{\langle \bar{K}^0|\mathcal{O}_{VV+AA}|K^0\rangle}{\frac{8}{3}\langle K^0|A_0|0\rangle\langle0|A_0|\bar{K}^0\rangle}\,,\label{eqn:bk}\end{equation}
where $|K^0\rangle$ is a neutral kaon state, $A_0$ is the zeroth component of the axial current and $\mathcal{O}_{VV+AA}$ is the $\Delta S = 2$ double weak decay operator responsible for the mixing, which has the form
\begin{equation}\mathcal{O}_{VV+AA} = (\bar{s}\gamma^\mu d)(\bar{s}\gamma^\mu d) + (\bar{s}\gamma^5\gamma^\mu d)(\bar{s}\gamma^5\gamma^\mu d)\,.\end{equation}
The Green's function $\langle\bar{K}^0(t_1)|\mathcal{O}_{\Gamma\Gamma}(t)|\bar{K}^0(t_2)\rangle$ has two Wick contractions
\begin{equation}
\mathcal{W}^1(t;\Gamma) \equiv 
\sum_{\vec{y}}\mathrm{tr}\;\Phi_1(\vec{y},t;\Gamma)\times \mathrm{tr}\;\Phi_2(\vec{y},t;\Gamma)
\end{equation}
and
\begin{equation}
\mathcal{W}^2(t;\Gamma) \equiv 
\sum_{\vec{y}}\mathrm{tr}\big{(}\Phi_1(\vec{y},t;\Gamma)\Phi_2(\vec{y},t;\Gamma)\big{)}\,,
\end{equation}
where
\begin{equation}
\Phi_i(\vec{y},t;\Gamma) \equiv \sum_{\vec{u},\vec{v}}\propf{s}{\fry{t}}{\fru{t_i}} \propfdag{d}{\frv{t_i}}{\fry{t}}\gamma^5\Gamma\,.
\label{equation:bkop}\end{equation}
Here, $\vec{y}$ is a vector on the interaction timeslice $t$; $\Gamma$ is the relevant product of gamma matrices $\Gamma\in\{\gamma^\mu,\gamma^5\gamma^\mu\}$; the subscript on the propagator refers to the quark flavour and the $\gamma^5$-Hermiticity of the propagator has been used as before.

RBC \& UKQCD typically use propagators calculated from a pair of spatially separated gauge fixed walls of Kronecker delta sources (referred to as \textit{GFWall} sources). These are usually calculated with both periodic ($p$) and antiperiodic ($a$) boundary conditions, using the $p+a$ combination to eliminate unwanted round-the-world contributions to the three-point function by doubling the periodicity of the meson's propagation. This method has been used to calculate $B_K$ on this $16^3\times 32$ ensemble \cite{Antonio:2007pb} using \textit{GFWall} sources on timeslices $t_1=5$ and $t_2=27$.

In this paper we use a variant of this method, using a single \textit{GFWall} source at time $\tau$ with any antiperiodic signs implemented on all time-directed links $U_t(\tau-1,\vec{x})$. Here we may take the $p+a$ combination as a forwards propagating solution and $p-a$ as a backwards propagating solution. This method has been used by Aubin et al. \cite{Aubin:2007pt} for the removal of round-the-world pion propagation in the calculation of the pseudoscalar decay constant.

\subsection{Stochastic $B_K$}

The form of equation (\ref{equation:bkop}) is suitable for calculation using both \textit{Z2PSWall} and \textit{Z2SEMWall} sources. However the stochastic cancellation to a delta function in space supresses the cross-terms by design. This property can be removed by choosing the same set of stochastic numbers on each site of the timeslice for a given hit, such that $M(\vec{x},\vec{y})$ becomes position independent and yet still retains the delta function in spin and colour space in the large hit limit. Thus, we introduce two new source types with this property: The \textit{Z2PSGFWall} and the \textit{Z2SEMGFWall} source which otherwise share the same source structure as the existing stochastic types. These sources should be used on gauge fixed configurations.

\subsection{$B_K$ results}

We calculated $B_K$ on the $16^3\times 32$ ensemble detailed in table \ref{table:ensembleparams}, using a valence quark mass of $0.04$ lattice units for all propagators. Fixing to Coulomb gauge, we used propagators generated from the \textit{Z2PSGFWall} and \textit{Z2SEMGFWall} stochastic source types and the standard \textit{GFWall} source. We also generated propagators from the \textit{Z2PSWall} and \textit{Z2SEMWall} types without gauge fixing. Following the discussion in section \ref{section:averaging} we generate a single stochastic hit per configuration for all source types bar the \textit{Z2PSWall} and  \textit{Z2PSGFWall}. For the latter types, the limited size of the ensemble forced us to increase the number of hits per configuration to four in order to compare with the \textit{GFWall} results at reasonable statistics. For a fair comparison at a given cost, the configurations used for the \textit{GFWall} correlators were spread across the ensemble in order to reduce the effect of autocorrelations. 

Upon loading each gauge configuration, we perform a spatial translation by a predermined four-vector $\vec d$. With every new configuration $\vec d$ is incremented by an amount $\vec \Delta$, allowing us to spread the sources throughout the lattice volume without the need to alter the location of the timeplane upon which we apply the boundary conditions: The sources are always placed at $t=0$ on the shifted configuration with the boundary conditions applied on the boundary between $t=T-1$ and $t=0$. We thus intended that the n$^{\mathrm{th}}$ configuration be shifted $\vec d _n = \vec d_1 + (n-1)\vec\Delta$, where the periodicity of the lattice is implicit. While this rule was mostly followed in contiguous segments, due to restarting the code this 
rule was interrupted at several points in the chain. The actual source origins are widely distributed and for the most part follow the above rule, and thus this will not substantially affect the conclusions. Our subsequent production running for phenomenological calculations using these methods follow the above rule strictly.

The fit ranges were chosen based upon the plateau range of the PA correlators in the denominator of equation (\ref{eqn:bk}), and the errors were estimated using the jackknife procedure. The data were binned over a minimum of 40 molecular dynamics time units as before.

\TABLE[p]{
\centering
\begin{tabular}{l|lllll}
\hline\hline
Source Type & \#conf. & Fit range & Fit value & $\chi^2 /\mathrm{d.o.f.}$ & Scaled fit\\
\hline
\textit{Z2PSWall} & 384 & 7-22 & 0.6338(75) & 1.882 & 0.6338(160)\\ 
                 &      & 9-24 & 0.6736(88) & 1.039 & 0.6736(188)\\
                 &      & 7-25 & 0.6374(83) & 1.185 & 0.6374(177)\\
                 &      &9-24 & 0.6479(68)  & 1.162 & 0.6479(145)\\
\hline
\textit{Z2PSGFWall} & 384 & 7-24 & 0.6155(79) & 0.721 & 0.6155(198)\\
                    &     & 9-23 & 0.6595(81) & 1.092 & 0.6595(203)\\
                   &      & 10-23 & 0.6492(94) & 0.857 & 0.6492(235)\\
                   &      & 8-22 & 0.6250(85) & 1.108 & 0.6250(213)\\
\hline
\textit{Z2SEMWall} & 128 & 9-24 & 0.6752(56) & 0.753 & 0.6752(173)\\
\textit{Z2SEMGFWall} & 128 & 10-24 & 0.6685(43) & 1.874 & 0.6685(67)\\
\hline\hline
\end{tabular}
\caption{Results for $B_K$ on the $16^3\times 32$ ensemble for the various source types calculated at a fixed cost of 384 inversions. The number of configurations is given in the second column. Four independent hits over the same set of configurations were calculated for the \textit{Z2PS} types; here we quote the results of independent fits to each of these sets in order to demonstrate the fluctuations in the correlators resulting from the choice of different random numbers. These data are inconsistent and thus we scale the errors by a PDG scale factor with the results given in the last column. PDG scale factors are calculated for the \textit{Z2SEM} types by splitting the available data into two sets and performing separate fits as discussed below.}
\label{table:bk384results}
}

\TABLE[p]{
\centering
\begin{tabular}{l|llll}
\hline\hline
Source Type & Fit range & Fit value & $\chi^2 /\mathrm{d.o.f.}$ & Scaled fit\\
\hline
\textit{Z2SEMWall} & 9-23 & 0.6626(39) & 1.010 & 0.6626(121)\\
                   & 8-23 & 0.6815(49) & 1.193 & 0.6815(152)\\
		   \hline
\textit{Z2SEMGFWall} & 8-24 & 0.6665(48) & 0.801 & 0.6665(75)\\
                     & 7 25 & 0.6572(38) & 1.138 & 0.6572(59)\\
		     \hline
\textit{GFWall}   & 7-25 & 0.6590(28) & 1.278\\
                     & 8-25 & 0.6579(24) & 1.081\\
\hline\hline		 
\end{tabular}
\caption{$B_K$ fits over 2 sets of 192 configurations with a separation of 10 configurations. The two sets are staggered by 5 configurations such that there is no overlap, thus approximating 2 hits on the same configurations.}
\label{table:bk192cfginterleaved}
}

\TABLE[p]{
\centering
\begin{tabular}{l|lllll}
\hline\hline
Source Type & \#conf. & Fit range & Fit value & $\chi^2/\mathrm{d.o.f.}$ & Scaled fit\\
\hline
\textit{Z2PSWall} & $4\times 384$ & 8-24 & 0.6548(51) & 0.261 & 0.6548(109)\\
\textit{Z2PSGFWall} & $4\times 384$& 8-25 & 0.6365(50) & 0.676 & 0.6365(125)\\
\textit{Z2SEMWall} & 384 & 7-23 & 0.6728(30) & 0.755 & 0.6728(93)\\
\textit{Z2SEMGFWall} & 384 & 7-24 & 0.6653(28) & 0.913 & 0.6653(43)\\
\textit{GFWall} & 128 & 9-23 & 0.6554(32) & 2.000\\
\hline\hline
\end{tabular}
\caption{Results for $B_K$ on the $16^3\times 32$ ensemble for the various source types. These data are calculated at a fixed cost of 1536 inversions, where the \textit{Z2PS} types were evaluated for 4 hits over the same 384 configurations. The value quoted in Antonio et al. \cite{Antonio:2007pb} is $0.659(3)$.}
\label{table:bkresults}
}
\clearpage

Table \ref{table:bk384results} shows the results for the stochastic types calculated at a fixed cost of 384 inversions. The \textit{GFWall} results have been omitted from this table due to the lack of statistics at this number of inversions (32 configurations). The fits to each of the 4 hits of \textit{Z2PSWall} and \textit{Z2PSGFWall} show discrepancies in their central values outside of the quoted errors, with a combined $\chi^2/\mathrm{d.o.f.}$ of $4.569$ and $6.274$ respectively. No improvement in the agreement was found by increasing the bin size. We therefore apply a PDG scale factor of $\sqrt{\chi^2/\mathrm{d.o.f.}}$ to the \textit{Z2PS} error bars in order to account for this unlikely, high $\chi^2/\mathrm{d.o.f.}$ The results of this scaling are given in the far right column.

In order to investigate the appropriate scaling factors for the other source types, we consider the fits to 2 sets of $192$ configurations with a separation of $10$ and a bin size of $40$ MD time units (4 configurations). The sets are staggered by 5 time units such that, due to correlations between nearby configurations, this method approximates two hits on the same $192$ configurations without the need for further computation. The results of this analysis are presented in table \ref{table:bk192cfginterleaved}. From each of these pairs of fits we calculate the combined $\chi^2/\mathrm{d.o.f.}$ allowing us to estimate the PDG scale factor as above. It is clear that the \textit{GFWall} results agree very well within errors, with a $\chi^2/\mathrm{d.o.f.}$ of $0.091$ and therefore need no scaling. However, the agreement between the two fits for the \textit{Z2SEM} correlators is poorer, with a $\chi^2/\mathrm{d.o.f.}$ of $9.591$ for the \textit{Z2SEMWall} results and $2.436$ for the \textit{Z2SEMGFWall}. Again we scale the \textit{Z2SEM} results by the PDG scale factor, with the results included in tables \ref{table:bk384results} and \ref{table:bk192cfginterleaved}.

Combining all available hits of the stochastic source types at a fixed cost of $1536$ inversions allows for their comparison with the \textit{GFWall} source correlators calculated on 128 configurations. These results are presented in table \ref{table:bkresults}, where we have scaled the stochastic source results by their appropriate PDG scale factors as before. After rescaling, all of the fits presented agree with the value quoted in Antonio et al. \cite{Antonio:2007pb} of  $0.659(3)$. From these data it is evident that the stochastic approach shows no advantage over the traditional method for the calculation of the $B_K$ matrix element, although the \textit{Z2SEMGFWall} correlators, which have the same structure as the \textit{GFWall} correlators in the large $N_{\mathrm{hits}}$ limit, give comparable results at the same cost.

\subsection{Comparison of the two-wall and single-wall approach to $B_K$}

\TABLE[t]{
\centering
\begin{tabular}{l|llll}
\hline\hline
Source Configuration & Fit range & Fit value & $\chi^2 /\mathrm{d.o.f.}$\\
\hline
1 wall, $t=0$ & 7-25 & 0.6591(28) & 1.353\\
\hline
2 walls, $t_1=5$, $t_2=27$ & 9-23 & 0.6634(26) & 0.779\\
\hline\hline
\end{tabular}
\caption{Fits to the $B_K$ three-point function using \textit{GFWall} sources at fixed cost, comparing the single-wall source method to the traditional two-wall source method.}
\label{table:bk1wallvs2}
}

\FIGURE[ht]{
\includegraphics[scale=0.4, angle=270]{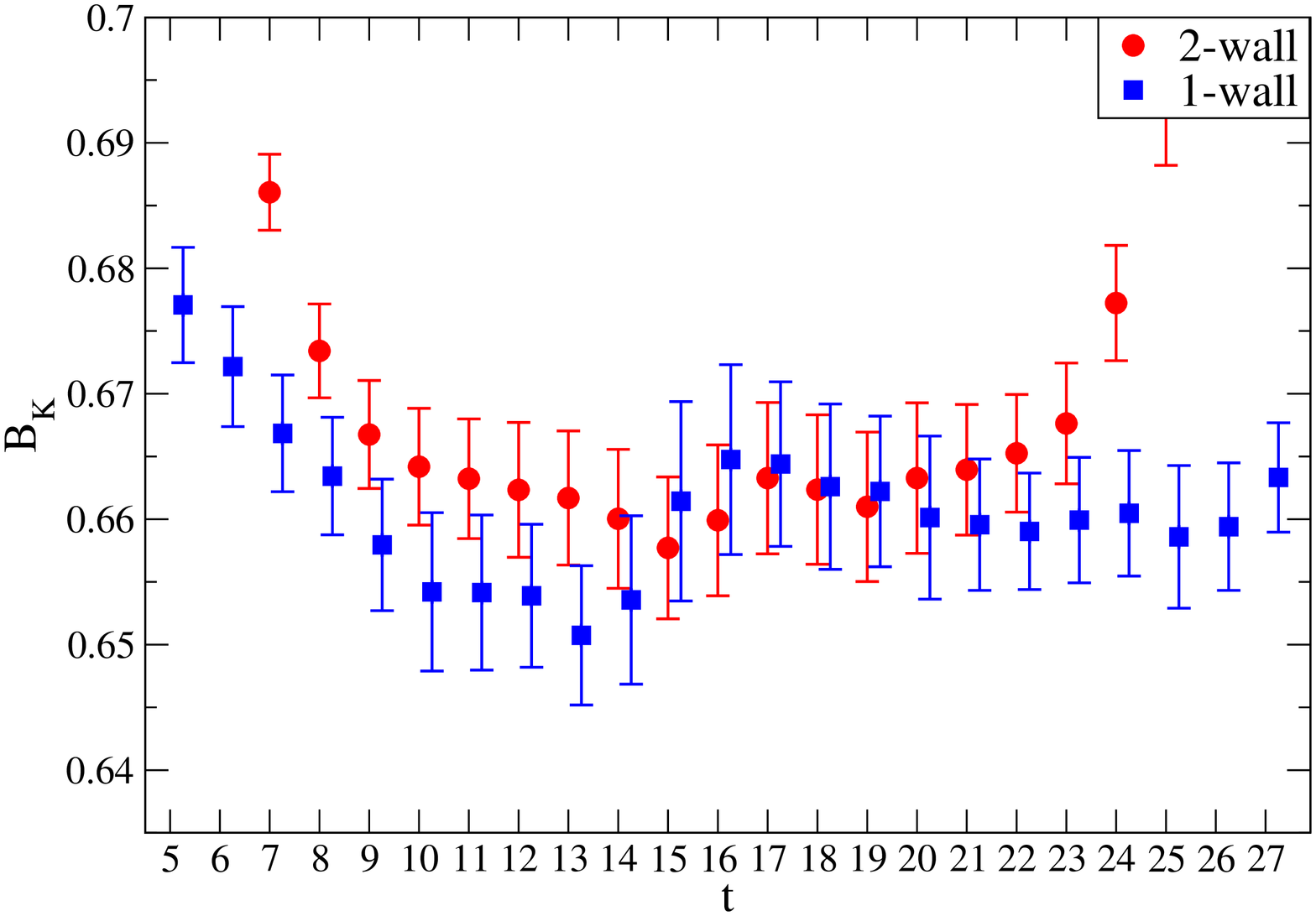}  
\caption{A comparison of the $B_K$ plateau calculated using the single-wall and two-wall approaches at a fixed cost in inversions. The two-wall sources reside on timeslices 5 and 27.}
\label{figure:bk1vswall2}
}

Using data from ref. \cite{Antonio:2007pb} we are able to compare the single-wall approach to the calculation of $B_K$ with the traditional two-wall method. Table \ref{table:bk1wallvs2} compares the fits to $B_K$ calculated with the two methods at fixed cost, using 196 configurations of single-wall data and 98 configurations of two-wall data. These sets of configurations overlap, and have configuration separations of 10 and 20 molecular dynamics time units respectively such that the set lengths are similar. Both sets of data were analysed with a bin size of 40 MD time units. The gauge fields of the single-wall calculation were shifted between configurations in order to reduce the effects of autocorrelations, whereas the two wall sources were fixed at $t=5$ and 27. Figure \ref{figure:bk1vswall2} compares the $B_K$ plateaus of the two methods. From these data we conclude that the single-wall method gives equivalent results at fixed cost. However, the method affords the sampling of more timeslices and configurations than the two-wall approach for the same cost.
\section{Stochastic calculation of three-point hadronic form factors}
\label{section:kl3}
The $\bar{K}^0\rightarrow \pi^+ l\nu_l$ form factor $K_{l3}$ and the pion electromagnetic form factor are phenomenologically interesting parameters calculated from relatively simple meson three-point functions and the meson two-point correlators discussed in section \ref{section:twopoint}.

Using the notation of Boyle et al. \cite{Boyle:2007wg}, the three-point functions have the form
\begin{eqnarray}
C_{P_iP_f}(t_{i},t,t_{f},\vec p_i,\vec p_f) &=&
\sum_{\vec{x}_f,\vec{x}} e^{i\vec{p}_f\cdot(\vec{x}_f-\vec{x})}
e^{i\vec{p}_i\cdot\vec{x}} \langle\, O_f(t_{f},\vec x_f)\,
V_\mu(t,\vec{x})\,O_i^\dagger(t_{i},\vec 0)\,\rangle\nonumber\\%[2mm]
    &=&
        \frac{Z_i\,Z_f}{4E_iE_f}\, \langle\,P_f(\vec{p}_f)\,|\,V_\mu(0)\,|\,
        P_i(\vec{p}_i)\,\rangle\,\nonumber\\ &&\hspace{-0.8in}
        \times\left\{\theta(t_f-t)\,e^{-E_i(t-t_i)-E_f(t_f-t)}\ -\
\theta(t-t_f)\,e^{-E_i(T+t_i-t)-E_f(t-t_f)}\right\}\,,\label{eq:threept}
\end{eqnarray}
where pseudoscalar ($i,f \in\{\pi,K\}$) initial states $P_i$ and final states $P_f$, with energies $E_i$ and $E_f$ respectively, are created by the interpolating operators $\mathcal{O}_{i,f} = \bar\psi_1 \gamma^5 \psi_2$ with fermions of the appropriate flavour. $V_\mu$ is the vector current operator with appropriate quark flavours to allow the transition, and $Z_f=Z_i^*=\langle\, 0\,|\,O_f(0,\vec 0)|\,P_f\,\rangle$. The source and sink timeplanes $t_i$ and $t_f$ are typically fixed, with a large time separation to remove the round-the-world contribution.

After Wick contraction, the three-point function becomes the trace over contracted propagators
\begin{equation}
\mathrm{tr}\left(\prop(\vec 0,t_i \leftarrow \vec{x}_f,t_f)\gamma^5\prop(\vec{x}_f,t_f \leftarrow \vec{x},t)\gamma^\mu \prop(\vec{x},t \leftarrow \vec 0, t_i)\gamma^5\right)\,.\label{eqn:kl3trace}
\end{equation}
These propagators can be determined from a single source point $(\vec 0,t_i)$ using a standard point source for the propagator $\prop(\vec{x},t \leftarrow \vec 0, t_i)$ and a sequential propagator \cite{Martinelli:1988rr}
\begin{equation}
 \prop(\vec 0,t_i \leftarrow \vec {p}_f,t_f \leftarrow \vec{x},t)
    =\sum_{\vec{x}_f}\gamma_5\left(\prop(\vec {x},t\leftarrow \vec{x}_f,t_f)
        \gamma^5 \prop(\vec{x}_f,t_f \leftarrow \vec 0,t_i)\,
        e^{-i\vec{p}_f\cdot\vec{x}_f}\right)^\dagger \gamma_5
\end{equation}
for the product $\prop(\vec 0,t_i \leftarrow \vec{x}_f,t_f)\gamma^5\prop(\vec{x}_f,t_f \leftarrow \vec{x},t)$, including a Fourier transform over $\vec x _f$ to momentum $\vec p _f$ at the sink timeplane $t_f$. The trace is then Fourier transformed over the vertex position $\vec{x}$ with the phase factor $e^{i(\vec{p}_i-\vec{p}_f)\cdot \vec{x}}$ to complete the three-point correlator.
 
For zero spatial momentum $\vec{p}_i$ at $t_i$, a stochastic wall source can be used in place of the traditional point source, giving an estimate of the spatial volume average. The stochastic averaging to Kronecker deltas will occur on the source timeplane, with the second leg of the sequential propagator inverted on the stochastic solution vector. We note that although our stochastic sources explicitly project to zero source momentum, partially twisted boundary conditions \cite{Boyle:2007wg,Sachrajda:2004mi} can be used in conjunction with this method to apply a residual momentum $\vec{p}_i$.

From equation (\ref{eqn:kl3trace}) we see that the propagators are contracted at the source without an intervening gamma matrix, and thus these three-point functions are suitable for calculation with \textit{Z2PSWall} sources, as well as the more general \textit{Z2SEMWall}.

This method has been adopted by the RBC \& UKQCD collaboration \cite{Boyle:2008yd} for the calculation of meson form factors. In the above, the authors compare the point source and stochastic approaches at a fixed statistical accuracy, concluding that the stochastic approach is vastly superior to the point source, offering similar statistical errors for less than ten percent of the computational cost. A similar approach is also used by the ETM collaboration \cite{Simula:2007fa}.

\section{Summary and conclusions}
In this paper we have detailed our investigations into the use of stochastic wall sources for the calculation of meson two-point and three-point functions using the one-end trick \cite{9810021,0603007}. We emphasise that this one-end method is a different application of stochastic sources to the method of approximating the all-to-all propagator for the calculation of disconnected correlation functions: The one-end trick uses the properties of the stochastic sources to offer a volume averaging of the standard connected correlation function alongside an overall reduction in computational cost.

In section \ref{section:stochsources} we have described in detail the structure of two $\zedtwocross$ stochastic wall source types, namely the \textit{Z2PSWall} and \textit{Z2SEMWall}, where the former is random in spin and colour space and the latter only in colour space, discussing the form of the noise introduced into various measurements. 

The viability of these source types for the calculation of meson two-point functions on the unit gauge and on a $16^3 \times 32$ Domain Wall QCD ensemble was demonstrated in sections \ref{section:unitgauge_meson} and \ref{section:results_meson}. We have shown that both stochastic source types give errors on the pseudoscalar meson mass that are smaller by a factor of two or more than those of the conventional point source approach at the same cost. In addition to the reduced error, there is a substantial improvement in the quality of the plateaus (figure \ref{figure:psmeffcost}) inspiring greater confidence in the results. 

In principle we believe that wall source techniques offer better sampling of low probability tails of the QCD functional distribution, for example physically significant contributions from rare, low eigenvalue modes
of the Dirac matrix. Such modes are likely to produce outliers when sampled by a point source. The relative improvement of wall sources over point sources will likely increase with lattice volume and 
decreasing quark mass, a feature in common with the low mode averaging approach.

The \textit{Z2SEMWall} is also shown to be viable for other meson correlators, showing improved statistical error over the point source for the vector meson mass. Thus we conclude that meson spectrum measurements such as masses and decay constants can be calculated with improved precision and confidence using the stochastic method.

In section \ref{section:stochasticbk} we have shown that stochastic sources are viable for the calculation of the kaon bag parameter, $B_K$. We have included two extra stochastic source types in the analysis that stochastically estimate the Coulomb gauge fixed wall source (\textit{GFWall}), each treating the spin-colour trace differently. These are referred to as \textit{Z2PSGFWall} and \textit{Z2SEMGFWall}. However, we found that the more complex structure of these three-point functions is less well treated by stochastic methods. Multiple measurements using stochastic sources on the same configurations showed disagreements in their central values outside of the jackknife error bars, forcing us to apply a PDG scale factor of $\sqrt{\chi^2/\mathrm{d.o.f.}}$ to the error bars of the stochastic results. We conclude that for three-point matrix elements of $\mathcal{O}_{VV+AA}$, the stochastic method offers no corresponding substantial gain over the traditional \textit{GFWall} method.

We also find that the use of a single \textit{GFWall} source calculated with periodic and antiperiodic boundary conditions, from which we calculate the forwards ($p+a$) and backwards ($p-a$) propagating components, offers comparable cost-effectiveness to the two-wall methods, but may allow more time origins or more configurations to be used when measurement cost is the limiting factor.

Finally, in section \ref{section:kl3} we discuss a method of stochastically estimating hadronic form factors. This method has been adopted by RBC\&UKQCD for a calculation of the $K_{l3}$ form factor \cite{Boyle:2008yd}, with the conclusion that a significant reduction in computational cost can be achieved for the same statistical error using the stochastic method coupled with partially twisted boundary conditions.

The relative difference in gain between the $K_{l3}$ form factor and $B_K$ can be explained as follows. The standard \textit{GFWall} method for the calculation of $B_K$ already provides an exact three-volume average of the operator insertion point, and thus the only benefit of the stochastic wall method is in the reduced cost of the spin-color tracing in the pseudoscalar interpolating operators. Our results suggest this is empirically ineffective.

In contrast the requirement of non-zero momentum for $K_{l3}$ results in a comparison of a localised source to a three-volume average, and there is much more scope for the stochastic volume average to gain. In this calculation, of course, momentum is injected using partially twisted boundary conditions, and cost of requiring multiple inversions for different momenta must be included.
It is certainly possible that, similar to $B_K$, a gauge fixed wall source in combination with partially twisted boundary conditions could result in an even greater improvement than the stochastic wall, by similarly providing an non-approximate volume average at similar cost. This is something we intend to study.

\acknowledgments

We thank Dong Chen, Saul Cohen, Calin Cristian, Zhihua Dong, Alan Gara, Andrew
Jackson, Chulwoo Jung, Changhoan Kim, Ludmila Levkova, Xiaodong Liao, Guofeng
Liu, Konstantin Petrov and Tilo Wettig for developing with us the QCDOC
machine and its software. This development and the resulting computer
equipment used in this calculation were funded by the U.S.\ DOE grant
DE-FG02-92ER40699, PPARC JIF grant PPA/J/S/1998/00756 and by RIKEN. This work
was supported by DOE grant DE-FG02-92ER40699 and PPARC grants
PPA/G/O/2002/00465 and PP/D000238/1. We thank RIKEN, BNL and the U.S.\ DOE for
providing the facilities essential for the completion of this work.

\appendix
\setcounter{figure}{0}%
\renewcommand{\thefigure}{A-\arabic{figure}}
\section{Binning analysis of the $16^3\times 32$ ensemble}  % use *-form to suppress numbering
\label{appendixa}

In order to prove any gain displayed in statistical errors is real and not associated with the spurious introduction of additional, correlated measurements we must ensure the independence of our extra data points. Thus we bin over adjacent configurations. The bin size was chosen by averaging over pseudoscalar meson correlators independently for each source type (\textit{Z2PSWall}, \textit{Z2SEMWall} and point) and fitting to a fixed range of $10-16$ while varying the bin size. We chose to include only a single \textit{Z2PSWall} hit on timeslice zero, as using all four available hits would weight the statistics in favour of this timeslice, making it more difficult to estimate the autocorrelations.
\FIGURE[t]{
\includegraphics[scale=0.4,angle=270]{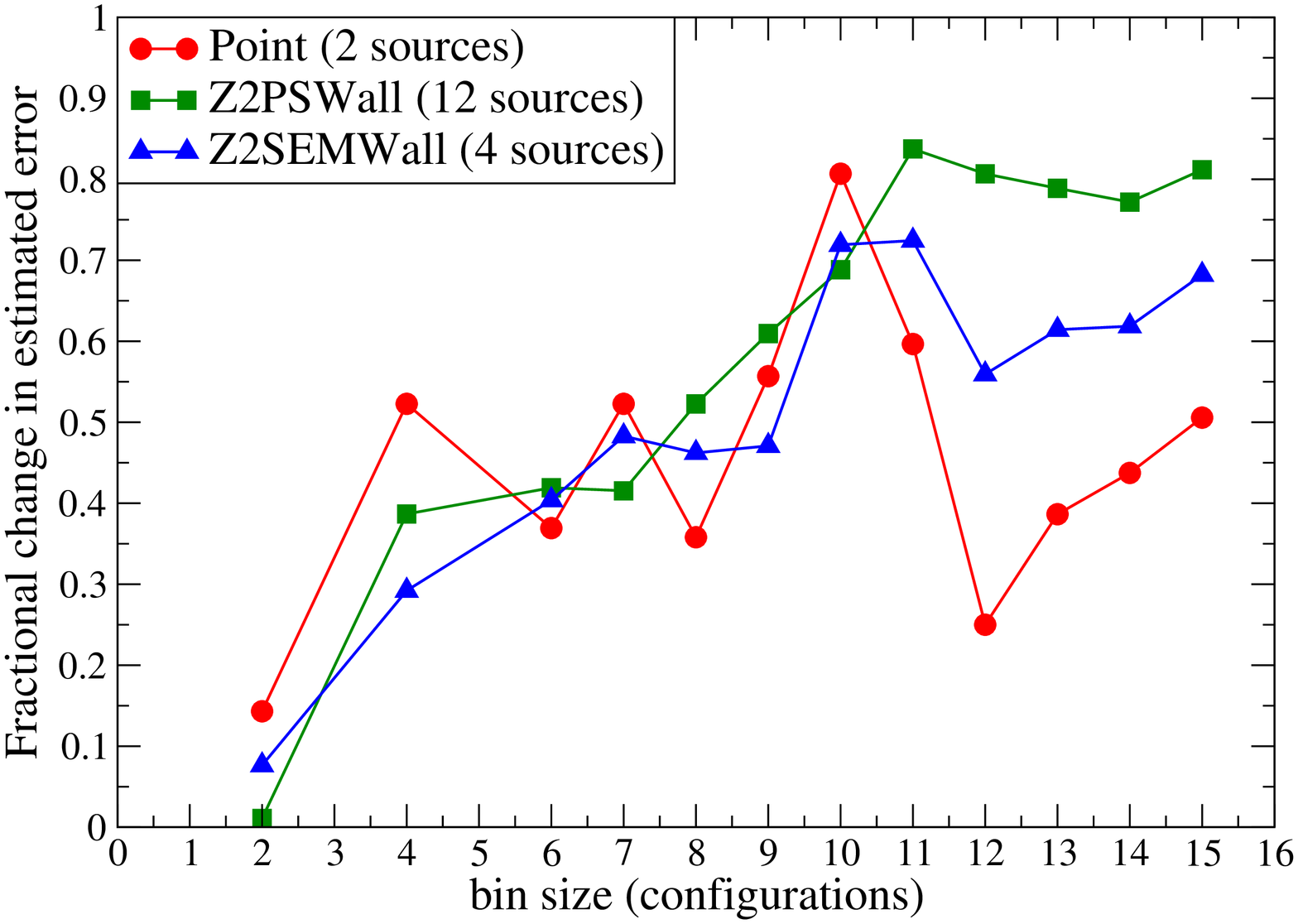}  
\caption{Fractional change in the estimated error on the fits to the source averaged pseudoscalar meson correlators, as a function of bin size. Fits are performed to the chosen fit range $10-16$.}
\label{figure:errbin}
}

Figure \ref{figure:errbin} shows the change in the estimated error on the fit as a function of the bin size for the chosen fit range, given as a fraction of the error on the unbinned data. The lack of smoothness of the curves in the figure indicates that the statistics are not good enough to provide anything but a rough estimate of the required bin size. The previous analysis \cite{ed-dwf-07}, using a combination of many source smearings and types,  placed a lower bound of 20 molecular dynamics time units on the integrated autocorrelation length of this ensemble, corresponding to a separation of 40 molecular dynamics time units (8 configurations) for independent measurements. This is in agreement with our analysis. Based upon this work and for comparison with the above we chose a final bin size of 8 configurations.

\end{document}